\documentclass[12pt,A4]{article}
\usepackage{graphicx}
\usepackage{epsfig}
\usepackage{epsf}
\begin{document}
\begin{center}
\large

{\bfseries 
MEASUREMENT OF THE EXTRACTED DEUTERON BEAM VECTOR POLARIZATION
AT NUCLOTRON}
\vskip 5mm
{L.S.Azhgirey$^a$}, 
M.Janek$^{a,b}$, A.N.Khrenov$^{a}$, V.P.Ladygin$^{a}$,
V.F.Peresedov$^{a}$, V.G.Perevozchikov$^{a}$, G.D.Stoletov$^{a}$,
 T.A.Vasiliev$^{a}$, V.N.Zhmyrov$^{a}$,  L.S.Zolin$^{a}$.
\vskip 5mm
{\small
$^a$ {\it
JINR, 141980, Dubna, Moscow region, Russia
}
\\
$^b$ {\it
University of P.-J. Shafarik, 041-54, Ko\v{s}ice, Slovakia
}
}
\end{center}
\vskip 5mm

\large

\begin{abstract}
The results of the measurements of the vector polarization of the 
extracted deuteron beam  at Nuclotron are presented. The intensity of 
the polarized deuteron beam during the measurements 
was $\sim 2.5\cdot 10^7$ particles/spill.
The measurements were made at the initial deuteron momenta of 5 and 
3.5 GeV/$c$. The averaged polarizations of the beam were
$ 0.606 \pm 0.014$ and $ 0.540\pm 0.019$, respectively.  

The investigation has been performed at LHE and LNP JINR.
\end{abstract}

\vskip 8mm

\newpage 

\large

\section{Introduction}

The experimental spin program proposed for Nuclotron (LHE-JINR)
requires good knowledge of the polarization
of the primary deuteron beam and/or a continuous monitoring
of the vector polarization stability during the experiment.
Such experiments are, for example,  $A{yy}$ \cite{ayy}, 
$PIKASO$ \cite{zolin},  
$PHe3$ \cite{khrenov}, 
$DELTA-SIGMA$ \cite{deltas}, 
$SMS-MGU$ \cite{ershov}, $STRELA$ and other, which 
are planned to be performed at Nuclotron. 

For these purposes the new version of the polarimeter \cite{gatchina,nim}
based on the measurement of the asymmetry of quasi-elasic
$pp$ scattering on hydrogen in $CH_2$ target has been installed at the  
focal point $F3$ in the LHE experimental hall.

The aim of this paper is to present the results of the 
vector polarization  measurements during first extraction of the
polarized deuteron  beam from Nuclotron in December 2002.

A brief description of the new version of on-line beam polarimeter is given in
Section 2.  
The results of the deuteron beam vector polarization 
measurements are given in 
Section 3. The effect of the trigger  is discussed in Section 4.
The conclusions are drawn in the last section.

\section{Polarimeter}

The measurements of the left-right 
asymmetry of quasi-elastic $pp$ scattering is the classical method 
to obtain the value of  nucleon beam polarization  
at intermediate energies. It has been used earlier  at
SATURNE II \cite{saclay}. The comparison of the elastic and quasi-elastic
$pp$ analyzing powers shows no difference between these quantities in a
very large energy range \cite{ball}. Using this
property, the LHE polarimeter \cite{gatchina,nim} measures 
$\vec P_B(d)$ of deuterons
provided by the ion source POLARIS \cite{polaris}.
The polarizations of protons and neutrons produced by the deuteron
breakup reaction in the forward direction are equal to each other; they
are related to the vector polarization of the deuteron beam.
%%%%%%%%%%%%%%%%%%%%%%%%%%%%%%%%%%%%%%%%%%%%%%%%%%%%%%%%%%
%%%%%%%%%%%%%%%%%%%%%%%%%%%%%%%%%%%%%%%%%%%%%%%%%%%%%%%
%   2   L.A.  The following seems to be unnecessary:
% , depending on the polarimeter acceptance. 
%%%%%%%%%%%%%%%%%%%%%%%%%%%%%%%%%%%%%%%%%%%%%%%%%%%%%%
%%%%%%%%%%%%%%%%%%%%%%%%%%%%%%%%%%%%%%%%%%%%%%%%%%%%%%%

The polarimeter for the measurement of the deuteron beam vector
polarization was installed close to the focal point $F3$ of the extraction beam line of
Nuclotron at
LHE, JINR. The details of the polarimeter were discussed in
ref.\cite{gatchina,nim}. Here we refer briefly the main changes 
made for new version of the polarimeter.

The layout of the polarimeter is given in Fig. 1.
Here $S_{1-12}$ are the scintillation counters, $IC$ is an ionization chamber,
$T$ is the target. 

The polarimeter measures the left-right (L-R) asymmetry of $pp$
quasi-elastic scattering, detecting both scattered and recoil particles
%%%%%%%%%%%%%%%%%%%%%%%%%%%%%%%%%%%%%%%%%%%%%%%%%%%%%%%
%%%%%%%%%%%%%%%%%%%%%%%%%%%%%%%%%%%%%%%%%%%%%%%%%%%%%
%   3   L.A.   Addendum: 
%   in coincidence. 
%%%%%%%%%%%%%%%%%%%%%%%%%%%%%%%%%%%%%%%%%%%%%%%%%%%%
%%%%%%%%%%%%%%%%%%%%%%%%%%%%%%%%%%%%%%%%%%%%%%%%%%%
in coincidence.
It consists of two pairs of arms in the horizontal plane installed at the
angles corresponding to $pp$~- elastic scattering kinematics. Each arm is
equipped by three scintillation counters. The six-fold (instead of four-fold in
the previous version \cite{gatchina, nim}) coincidence of
counter signals from each pair of conjugated
arms $S_1$ to $S_6$ and $S_7$ to $S_{12}$ define L or R scattering
events, respectively. 
The increasing of the level of coincidence reduces significantly the number of
random coincidences.
Also the ionization chamber $IC$ used as beam intensity monitor
is installed just in front of the polarimeter target $T$.

The sizes of plastic scintillators and the distance of the counters 
$S_1-S_{12}$ from the target point are given in Table 1.
The solid angles of the  forward and recoil arms are defined 
by the sizes and positions of the $S_1$, $S_8$ and $S_5$, $S_{11}$
counters, respectively.  The solid angle for $pp$~- elastic scattering
is defined by the acceptances of the forward arms,
while the recoil arms have larger acceptance.
The size of the forward arms determining counters ($S_1$ and $S_8$)
is $40\times 40$ mm$^2$ and their distance
from the target center is 1720 mm. 
Therefore the scattering angle acceptance and 
solid angle subtended by the L or R counters are 
$\Delta\theta =\pm 0.67^\circ$ and
$\Delta\Omega = 5.4\cdot 10^{-4}$~sr, respectively. 
The solid angle for the recoil particles is 
$\Delta\Omega_{rec} = 9\cdot 10^{-3}$~sr. Such an angle allows to
detect the recoil protons from $pp$~- elastic scattering without losses of
the statistics 
and with insignificant magnification of the admixture of the 
quasi-elastic events from the carbon content of $CH_2$ target.

Coincidence counts of the
polarimeter arms and monitor informations were recorded for each beam
polarization direction and stored by a PC data acquisition system after
the end of each beam spill.

The polarization of the extracted beam was oriented along the vertical
axis (perpendicularly to the beam momentum direction) and flipped
every accelerator spill.

The method of the vector polarization measurement is based on the 
detection of the particles scattered leftwards and rightwards.
The left-right asymmetry for a certain sign of the beam polarization
($\pm$) can be calculated from the relation
\begin{equation}
\epsilon^{\pm} = \frac{n_L^\pm/n_R^\pm - n_L^0/n_R^0}
{n_L^\pm/n_R^\pm + n_L^0/n_R^0},
\end{equation}
where $n^{\pm,0}_L$ and $n^{\pm,0}_R$ are the respective numbers 
of events scattered leftwards  and
rightwards for different spin states
of polarization source normalized to the beam intensity.

If an effective analyzing power of the polarimeter $A$ is known, 
the beam polarization $P^{\pm}$ can be calculated according to
\begin{equation}
P^{\pm} =\epsilon^{\pm}/A.
\end{equation}

\section{Measurements of the beam polarization}

The polarized deuterons were produced by the ion source POLARIS 
\cite{polaris}. The extraction of the polarized deuteron beam from 
Nuclotron 
has been performed at 5.0 GeV/c and 3.5 GeV/c. The intensity of 
the beam was measured by the ionization chamber $IC$ placed in 
front of the polarimeter. The results of the intensity 
measurements versus time are shown in Fig.2.

The averaged beam 
intensity was only $\sim 2-3\cdot 10^7$ particles per burst. 
Therefore, to have a reasonable counting rate for both left and 
right arms of the polarimeter the $CH_2$ target thickness was 
increased up to 5~cm, and the level of coicidences was decreased 
from 6 to 4 (or 3). 

The results of the asymmetry measurements for the both signs of the beam 
polarization are shown in Fig.3. The measurements were made at the 
initial deuteron momenta 5.0 GeV/$c$ and 3.5 GeV/$c$ (last 3 points in 
Fig.3).The forward scattering angle was set $14^\circ$ for the
both momenta, while the angle for the recoil proton was set in 
accordance with the kinematics of elastic $pp$ scattering. 
At higher momentum the absolute value of the asymmetry is lower 
due to a fall in effective analyzing power of the polarimeter 
versus energy \cite{gatchina,nim}. 

Since the polarimeter was not calibrated at these both deuteron momenta,
the values of the effective analyzing power $A(CH_2)$ were taken 
from the parametrization. 
The results of the linear and quadratic proton momentum dependences of the
analyzing power $A(CH_2)$
are presented in Fig.4 by the dashed and solid lines,
respectively.

The linear dependence of analyzing power on the  proton momenta
has the following form \cite{calib0}
\begin{eqnarray}
\label{lin}
A(CH_2)(p_p) &=& 0.6429 - 0.1628\cdot p_p,
\end{eqnarray}
while the quadratic dependence  is given as
\begin{eqnarray}
\label{quad}
A(CH_2)(p_p) &=& 0.5190 - 0.0456\cdot p_p - 0.0262\cdot p_p^2
\end{eqnarray}
The values of the  
effective  analyzing power $A(CH_2)$  
at the proton momenta 1.75 GeV/c and 2.5 GeV/c were taken 
according relation (\ref{quad}) as $0.359$ and $0.241$,  respectively.
Note, that the values of analyzing power $A(CH_2)$ at 1.75 GeV/c obtained 
from expressions (\ref{lin}) and (\ref{quad}) agrees with the precision
better than 0.5\%.

The values of the deuteron beam vector polarization at the  both 
deuteron momenta are presented in Table 2. 
The averaged over spin states values of polarization are 
$0.540 \pm 0.019$ and 
$0.606 \pm 0.014$ at 3.5 GeV/$c$ and 5.0 GeV/$c$, respectively.
Some difference in the polarization values at the both momenta can be due to
uncertanty in the values of effective analyzing power $A(CH_2)$ used.

%To calculate the beam 
%polarizations, the following values of the effective analyzing
%power were taken: $A(CH_2) = 0.235$ for $T_p = 1.73$ GeV and 
%$A(CH_2) = 0.357$ for $T_p = 1.05$GeV, to give
%$P^+ = 0.649 \pm 0.019,\, P^- = -0.593 \pm 0.020$ for 5 GeV/$c$ and
%$P^+ = 0.534 \pm 0.026,\, P^- = -0.551 \pm 0.027$ for 3.5 GeV/$c$ of 
%initial deuteron momentum (only statistical errors are given).
%\vspace{0.5cm}

\section{Trigger effect}

The level of coincidences of scintillation counter signal was 
reduced because of a low intensity of the beam during polarization 
measurements. In this case the effective solid angle of the polarimeter 
changes due to finite size of the beam and target.
 Therefore, the effective analyzing power
of the polarimeter could also change due to possible different yield 
from carbon content of the target.

The special study to test such an effect was
performed using unpolarized deuteron beam with the 
momentum of 3.5 GeV/c at Nuclotron run in June 2003.

The idea of these studies is based on the following
assumption. 
Let us suppose that an effective analyzing power of polarimeter $A$ 
may be represented as
\begin{equation}
 A = (1-k)\cdot A_{pp}+ k\cdot A_{pC},
\end{equation} 
where $A_{pp}$ and $A_{pC}$ are the analyzing powers of
$pp$ elastic scattering and $pC\to ppX$ reaction, respectively, and
$k$ is a coefficient proportional to the carbon content of $CH_2$ 
target. 
If the effective solid angle changes, the fraction of events from 
carbon also may change, 
and new value of an effective analyzing power $\bar{A}$ can be
written as:
\begin{equation}
 \bar{A} = (1-\bar{k})\cdot A_{pp}+ \bar{k}\cdot A_{pC},
\end{equation} 
where $\bar{k}$ is some new coefficient. 
%\vspace{0.5cm}
If $A_{pp},\ A, k$ and $\bar{k}$ are known, a corrected  value 
of the effective analyzing power $\bar{A}$ may be found from
\begin{equation}
 \bar{A} = \frac{\bar{k}}{k} \cdot A + \large ( 1-\frac{\bar{k}}{k}
 \large ) \cdot A_{pp}.
\end{equation} 

The value of $A_{pp} = 0.418$ is known from the fit to the world 
$pp$-data for the scattering at an angle of 14$^\circ$ 
(see Fig. 5), 
$A = 0.359$ can be taken from the previous calibration of polarimeter 
at the deuteron momentum of 3.5 GeV/$c$, and $k$ and $\bar{k}$ 
have been obtained from direct measurements in a special run at Nuclotron 
in June 2003.

The results of these measurements are presented in Fig.6
for 2 configurations of the polarimeter trigger: Trigger 1 
is 6-fold coincidences, while Trigger 2 is 4(3)-fold 
coincidences. The counting rates normalized for the beam 
intensity and target thickness for the $CH_2$ (open symbols) 
and carbon (filled symbols) are shown versus recoil
particle scattering angle. The solid lines are the results of the
fit of the $CH_2$ and carbon yield.

Since the carbon content, in general,  can be different for the 
left and right arms, both of them were considered as the separate 
polarimeters with their own effective analyzing powers.
The polarization of the beam was  calculated   
for the left and right arms of the polarimeter and 
weighted averaged for each sign of the polarized ion source.

Two methods to estimate possible trigger effect were used.
The first method (Method~1) is based on the subtraction of the direct 
measurements of
the event rates from $CH_2$ and carbon targets normalized to the 
beam intensity and
number of nuclei in the targets.
The second one (Method~2) uses the values obtained from the parameters 
of the fit for the $CH_2$ and carbon.
The results are given in Table 3.  
The first line contain the polarization values obtained without corrections,
while the second and the third ones give the values obtained
with the corrections using Method 1 and Method 2, respectively. 
It is seen that the values of polarization corrected with our methods 
agree in limits of error bars with those obtained initially.

Fig.7 demonstrates the normalized yield of the events from the 
hydrogen content of the 
$CH_2$ target obtained by the $CH_2-C$ subtraction.  
The cleaness of the subtraction, especially, for the 6-fold coincidences allows
to obtain  the values of the polarization by the use the data on the
analyzing power of $pp$ elastic scattering only, as it is proposed 
in ref. \cite{calib0}.
Such a procedure can be applied in future on polarized deuteron beam 
at Nuclotron.

\section{Conclusions}

The results of this work can be summarized as following.

The intensity of the firstly extracted polarized deuteron beam 
at Nuclotron during the measurements was $\sim 2.5\cdot 10^7$ particles/spill.

The polarimeter placed at focal point $F3$ measured significant
asymmetry at 5 GeV/c and 3.5 GeV/c for the vector polarized
deuteron beam. The polarization of the extracted deuteron beam averaged 
over the spin states was 
$ 0.606 \pm 0.014$ and 
$ 0.540\pm 0.019$ at 5 GeV/$c$ and 3.5 GeV/$c$, respectively.

Special study was performed to estimate the possible systematics
due to trigger effect using unpolarized deuteron beam. It was shown that 
the modification of the polarization values at 3.5 GeV/c due to this effect 
is small and does not exceed the statistical and systematic errors.

\vspace{0.5cm}
Authors thank the polarized ion source POLARIS  and  Nuclotron
accelerator staff for providing good conditions for the work.
This investigation has been supported in part by the Russian 
Foundation for Fundamental Research (grants $N^o$ 03-02-16224  
and $N^o$ 04-02-17107) and by the Grant Agency for
Science at the Ministry of Education of the Slovak Republic (grant
No.1/1020/04)

\newpage

\centerline{Figures caption}

{\bf Fig.1.} New version of the beam polarimeter at focal 
point $F3$. $IC$ is the ionization chamber, $T$ is the $CH_2$ 
(or $C$) target, $S_1-S_{12}$ are the scintillation counters.

{\bf Fig.2.} Intensity of the extracted polarized 
deuteron beam at Nuclotron in December 2002 run versus time.

{\bf Fig.3.} 
Asymmetry of  the extracted polarized 
deuteron beam at Nuclotron in 2002 December run at 5.0 and 
3.5 GeV/c (last 3 points) versus time.

{\bf Fig.4.} The fit of the energy dependence of the 
effective analyzing power $A(CH_2)$ 
at $14^\circ$ \cite{nim, calib0}.
The dashed and solid lines are the results of the 
parametrization by the linear and quadratic dependences on the
proton momentum, respectively.

{\bf Fig.5.} The fit of the energy dependence of the $A_{pp}$ 
analyzing power at $14^\circ$ \cite{calib0}.

{\bf Fig.6.} The normalized counting rate for the $CH_2$ 
(open symbols) and carbon (filled symbols) targets for different 
configurations of the trigger versus recoil
particle scattering angle. Figures a) and b) correspond to the 6 fold coincidences
for the left and right arms, while figures c) and d) 
correspond to the 4(3) fold coincidences, respectively.

{\bf Fig.7.} The normalized counting rate for the hydrogen content of the $CH_2$ 
target for different 
configurations of the trigger versus recoil
particle scattering angle. Figures a) and b) correspond to the 6 fold coincidences
for the left and right arms, while figures c) and d) 
correspond to the 4(3) fold coincidences, respectively.

 \newpage

{\bf Table~1.} Dimensions of plastic scintillators 
($x\times y\times z$), where $x$ and $y$ are the sizes in the horizontal
and vertical planes, respectively, and $z$ is the thickness) and their
distance from the target center. 
\vspace{1cm}

\begin{center}
\begin{tabular}{|c|c|c|c|}
\hline
Arm &Counter & Dimensions, & Distance from \\
    &    & $mm^3$       &  the target, $mm$ \\         
\hline
Forward &$S_1$, $S_8$ & $40\times40\times5$ & 1720 \\
        &$S_3$, $S_9$ & $40\times40\times5$ & 1260 \\
        &$S_2$, $S_7$ & $35\times35\times5$ & 835 \\
\hline
Recoil & $S_5$, $S_{11}$ & $50\times160\times8$  & 940 \\
       & $S_6$, $S_{12}$ & $45\times145\times8$  & 690\\
       & $S_4$, $S_{10}$ & $40\times130\times8$  & 460\\
\hline
\end{tabular}

\end{center}

\newpage

{\bf Table 2.} Vector polarization of
the extracted deuteron beam at incident momenta
3.5 and 5.0 GeV/$c$.

\begin{center}
\begin{tabular}{|c|c|c|c|}
\hline
$P_d$, GeV/$c$ & $P^+ \pm\Delta P^+$  & $P^- \pm\Delta P^-$ & $P \pm\Delta P$ \\
\hline
3.5 &  $0.531 \pm 0.026$ &  $-0.548 \pm 0.027$ &  $0.540 \pm 0.019$ \\ 
5.0 &  $0.633 \pm 0.019$ &  $-0.578 \pm 0.020$ &  $0.606 \pm 0.014$ \\
\hline
\end{tabular}

\end{center}

\newpage

%{\small
{\bf Table 3.} The beam polarization for the both spin states of 
the polarized ion source without and with the correction for the 
trigger effect.
%}
%\vspace{0.5cm}

\begin{center}

\begin{tabular}{|c|c|c|c|}
\hline
 Method & $P^+\pm \Delta P^+$ & $P^-\pm \Delta P^-$ & $P\pm \Delta P$ \\
\hline
Without   & $0.531\pm   0.026$ & $ -0.552\pm   0.027$ 
& $0.541\pm   0.019$ \\
corrections  & & & \\
Method 1 & $0.540\pm   0.027$ & $ -0.552\pm   0.027$
&  $0.546\pm   0.019$ \\
Method 2  & $0.563\pm   0.028$ & $ -0.564\pm   0.028$
&  $0.563\pm   0.020$ \\
\hline
\end{tabular}
\end{center}

\newpage

\begin{figure}[hbtp]
    \resizebox{16cm}{!}{\includegraphics{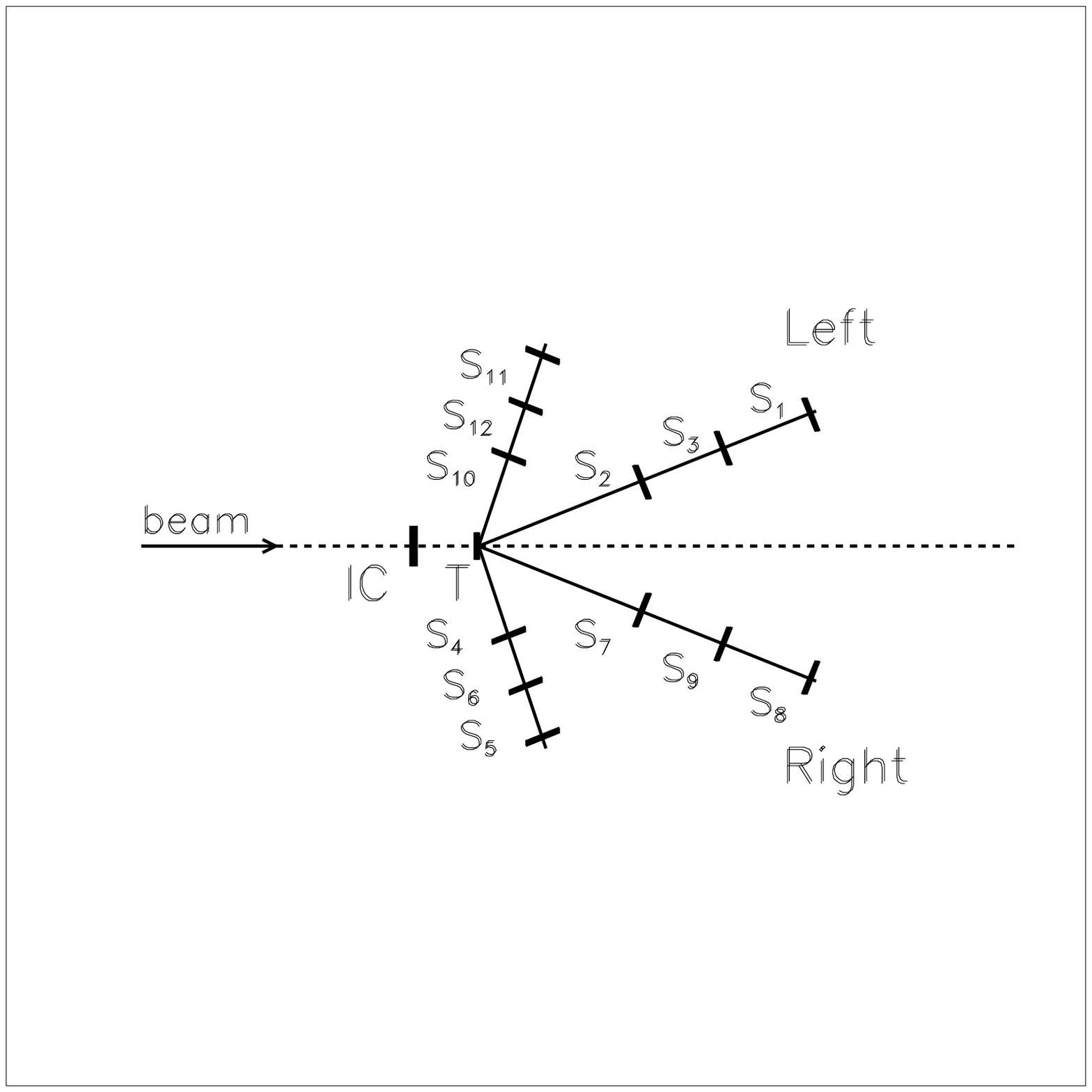}}
  \label{fig:polarim}
\end{figure}

~~
\vspace{2cm}
\centerline{\bf Fig.1}

\newpage

\begin{figure}[hbtp]
    \resizebox{16cm}{!}{\includegraphics{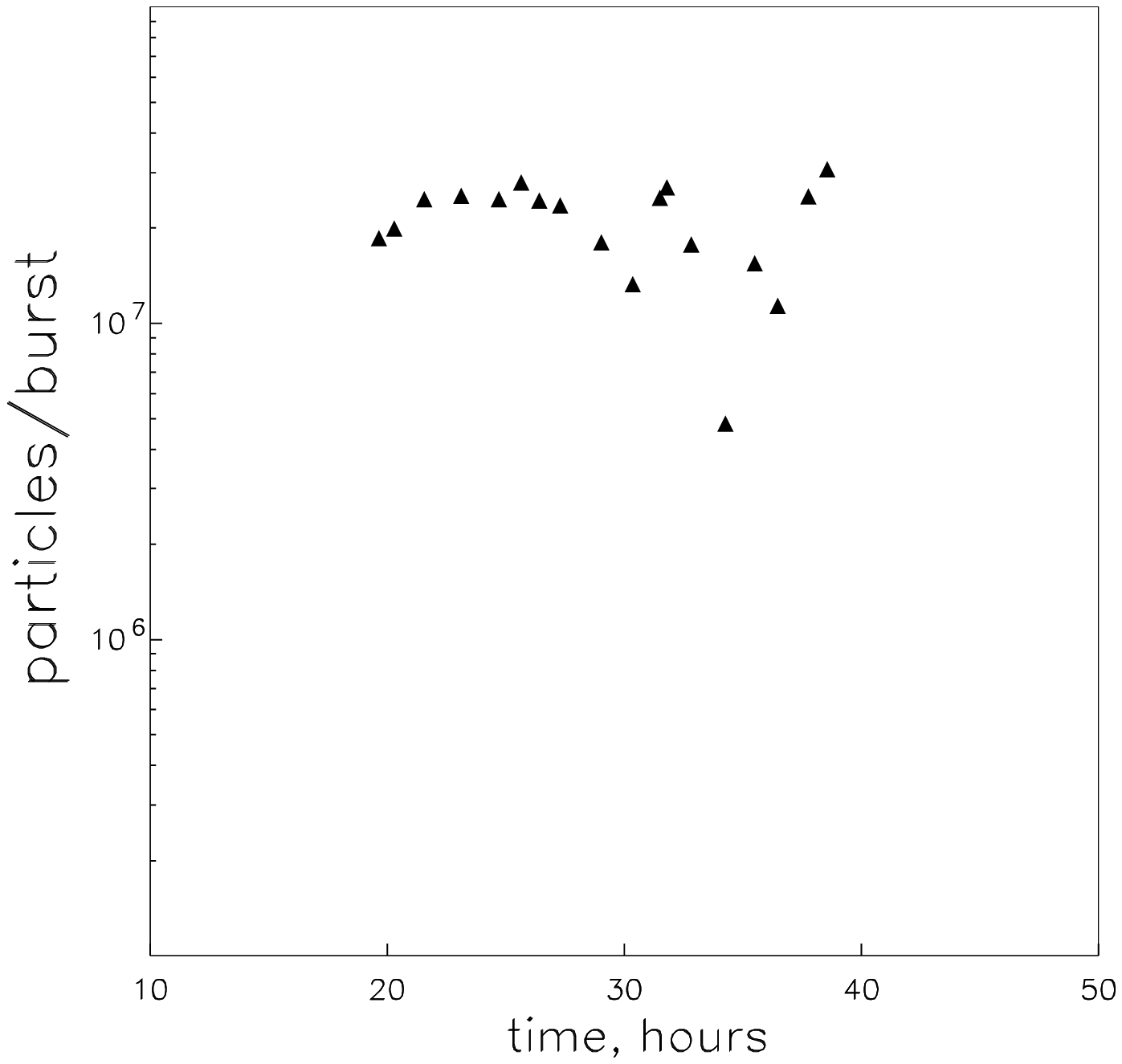}}
  \label{fig:polarim}
\end{figure}

~~
\vspace{2cm}
\centerline{\bf Fig.2}

\newpage

\begin{figure}[hbtp]
    \resizebox{16cm}{!}{\includegraphics{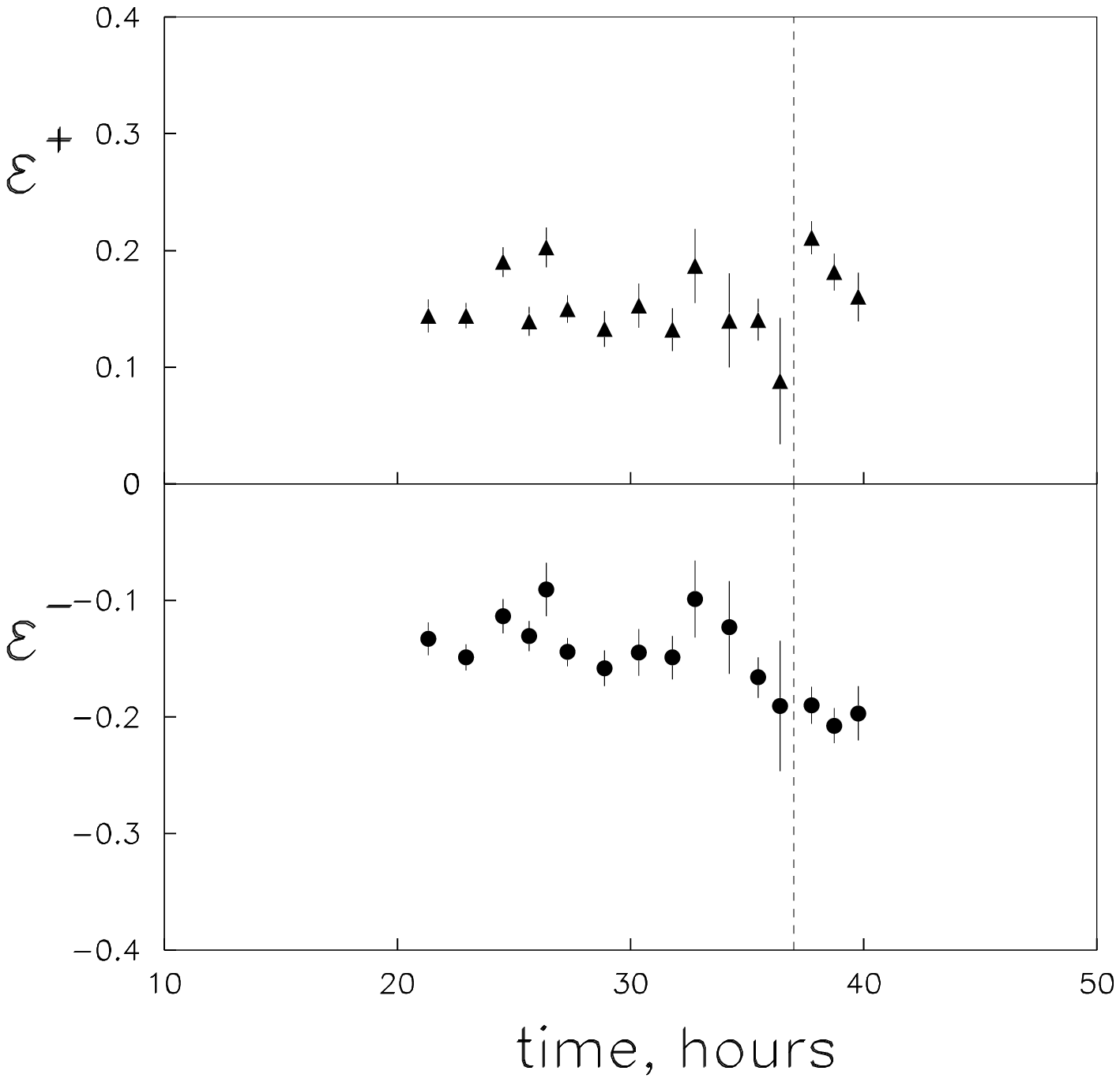}}
  \label{fig:polarim}
\end{figure}

~~
\vspace{2cm}
\centerline{\bf Fig.3}

\newpage

\begin{figure}[hbtp]
    \resizebox{12cm}{!}{\includegraphics{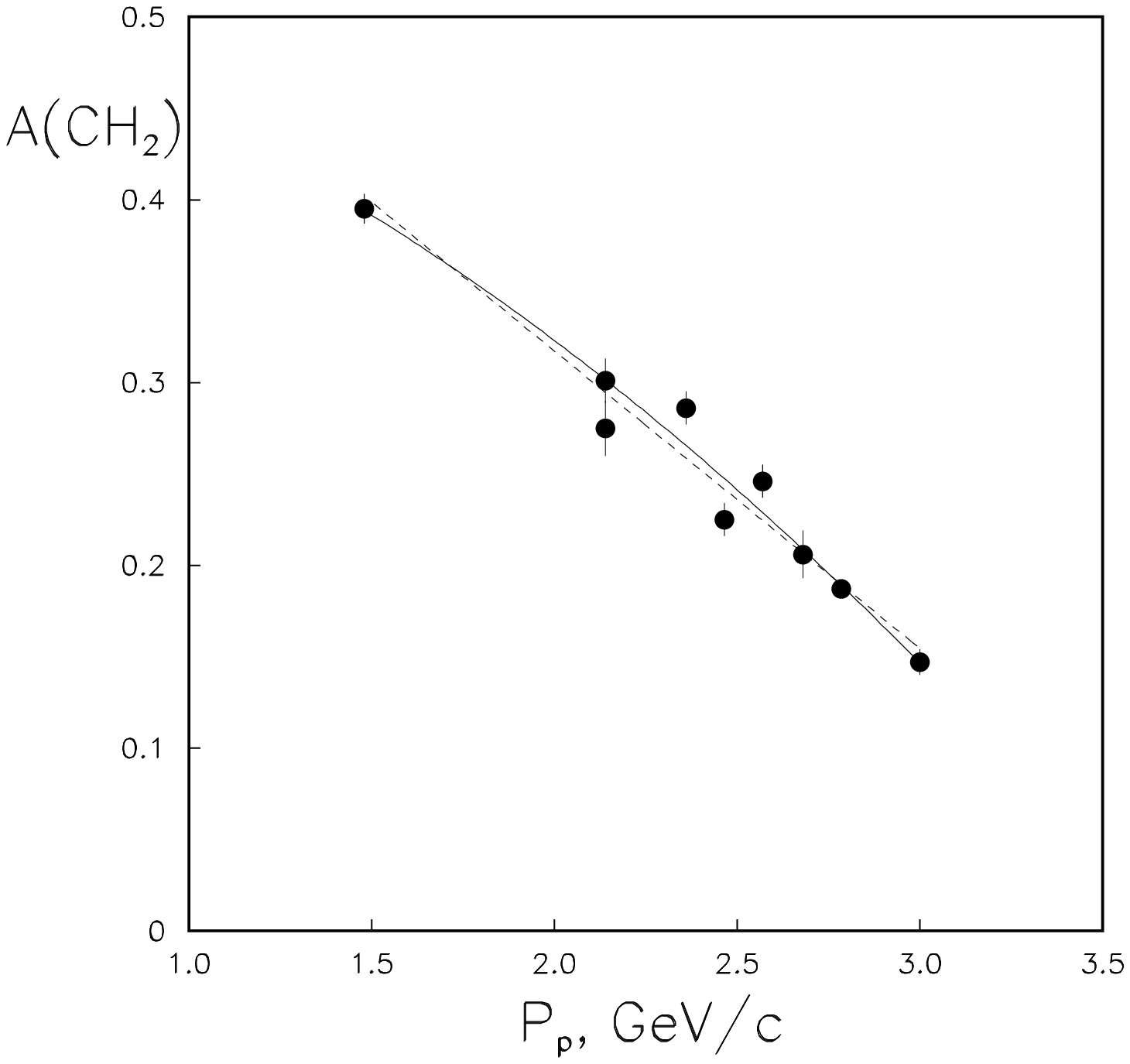}}
  \label{fig:polarim}
\end{figure}

%~~
%\vspace{2cm}
\centerline{\bf Fig.4}

\newpage

\begin{figure}[hbtp]
    \resizebox{12cm}{!}{\includegraphics{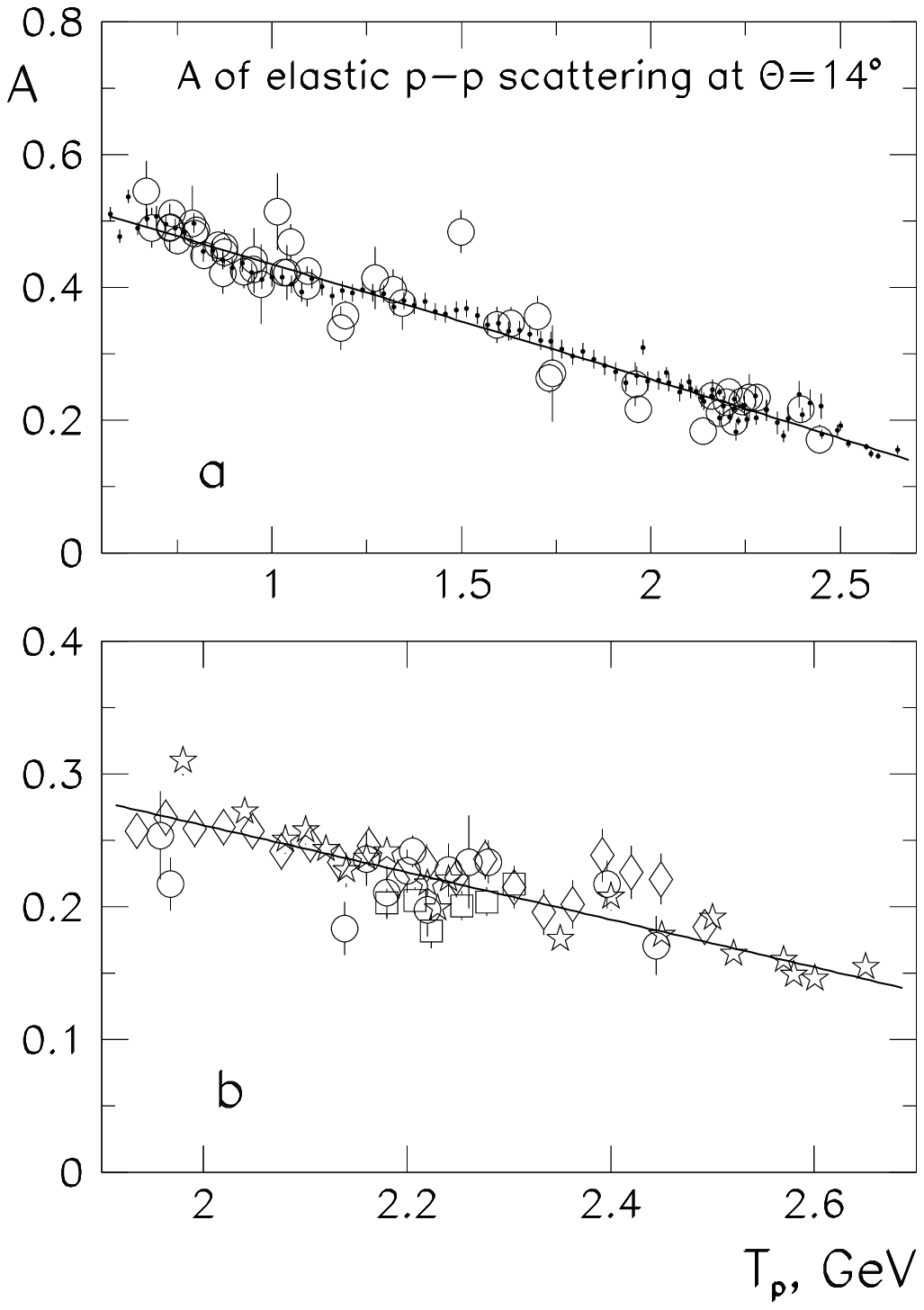}}
  \label{fig:polarim}
\end{figure}

~~
\vspace{2cm}
\centerline{\bf Fig.5}

\newpage

\begin{figure}[hbtp]
    \resizebox{16cm}{!}{\includegraphics{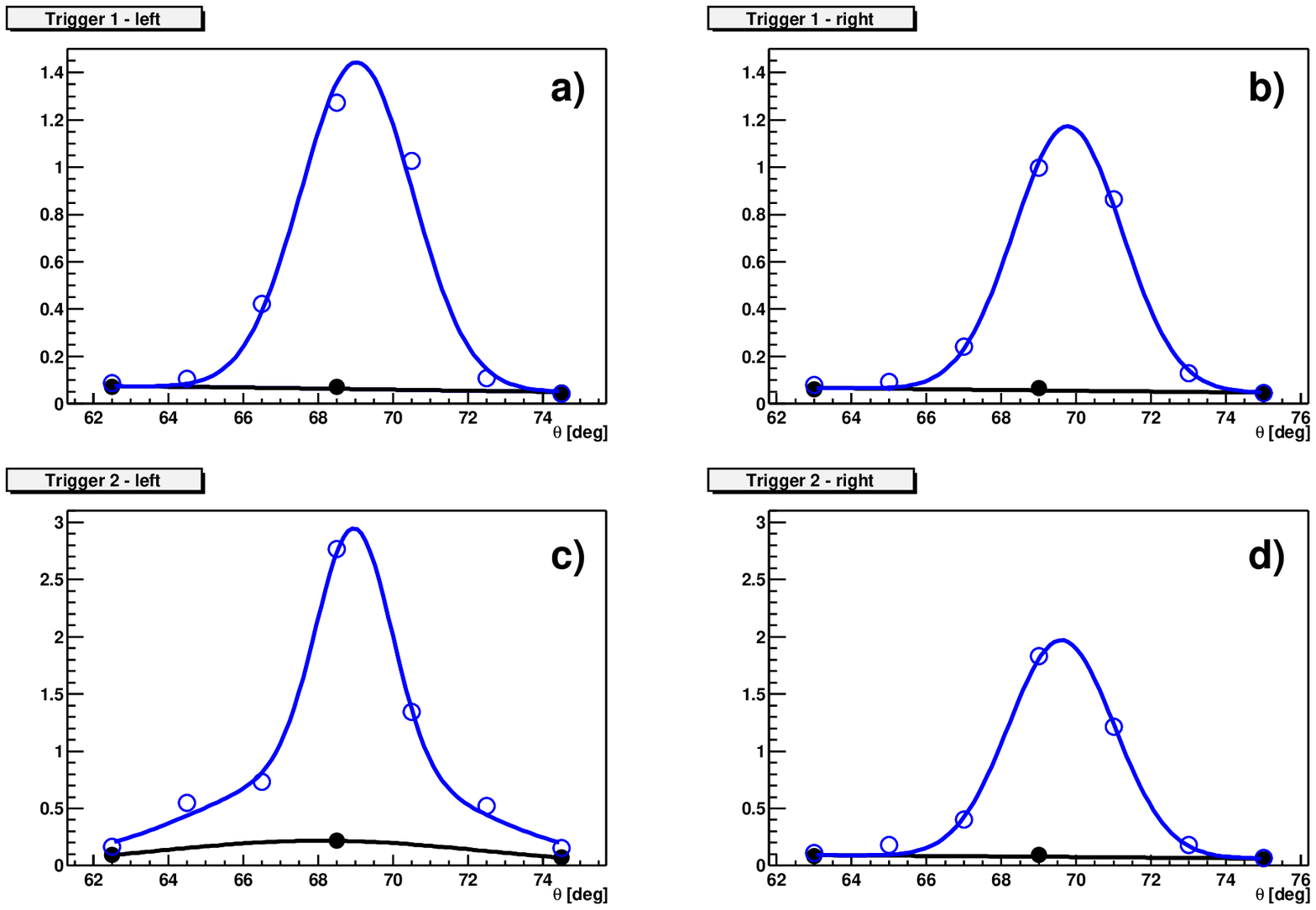}}
  \label{fig:trig_f1}
\end{figure}

~~
\vspace{2cm}
\centerline{\bf Fig.6}

\newpage

\begin{figure}[hbtp]
    \resizebox{16cm}{!}{\includegraphics{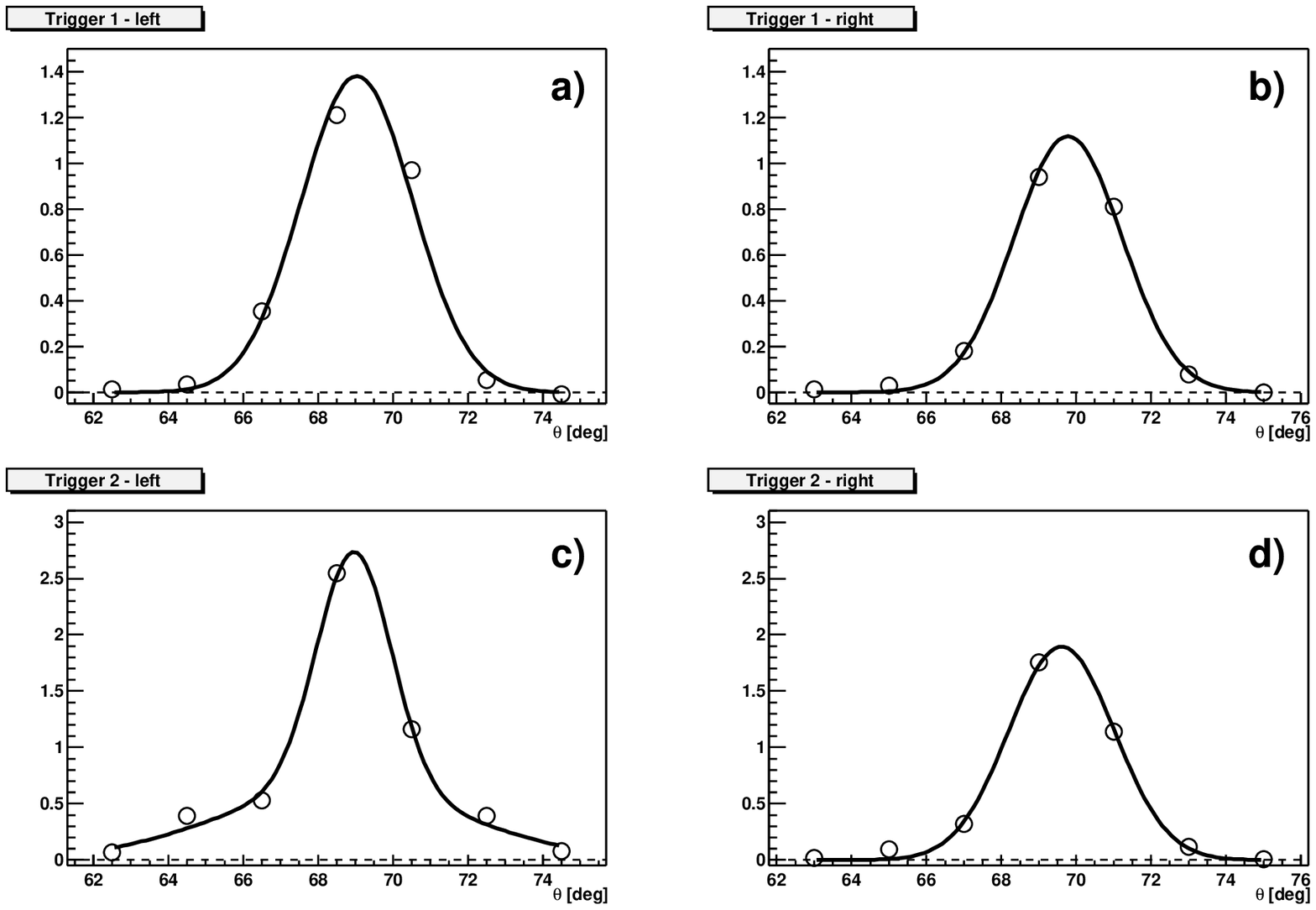}}
  \label{fig:trig_f2}
\end{figure}

~~~~

\vspace{2cm}
\centerline{\bf Fig.7}

\end{document}